\documentclass[11pt]{article}
\usepackage{epsfig}
\newcommand{\mubest}{\mu_\mathrm{best}}

\newcommand{\nexp}{n_0}
\begin{document}
\vspace*{-2cm}
\noindent
\hspace*{11cm}
UG--FT--108/99 \\
\hspace*{11cm}
hep-ex/9911024 \\
\hspace*{11cm}
November 1999 \\
\vspace{2cm}
\begin{center}
\begin{large}
{\bf Computation of confidence intervals for Poisson processes}
\end{large}

\vspace{0.5cm}
J. A. Aguilar--Saavedra \footnote{e-mail address: {\tt
 aguilarj@ugr.es}} \\
{\it Departamento de F\'{\i}sica Te\'{o}rica y del Cosmos \\
Universidad de Granada \\
E-18071 Granada, Spain}
\end{center}
\begin{abstract}
We present an algorithm which allows a fast numerical computation of
Feldman-Cousins confidence intervals for Poisson processes, even when the
number of background events is relatively large. This algorithm incorporates an
appropriate treatment of the singularities that arise as a consequence of
the discreteness of the variable.
\end{abstract}
\hspace{0.8cm} \noindent
PACS: 02.70.-c, 06.20.Dk, 29.85.+c

\vspace{0.5cm}
\section{Introduction}
In physics there are many situations where the outcome of an experiment is a
positive integer number with a Poisson distribution. This is the case for
instance of the number of events of a certain type produced in high energy
collisions. The statistical analysis of these processes is a difficult task
when the result obtained is in the limit of the sensitivity of the experiment. 
In general, the number of events $n_0$ obtained in an experiment consists of
background events with known mean $b$ and signal events, whose mean $\mu$ is the
quantity that we want to determine. The problem arises when the number of events
obtained $n_0$ is significantly lower than the background expected $b$.
This happens in some experiments on neutrino oscillations, for instance in the
KARMEN 2 experiment \cite{papiro1}. 

Usually, after performing an experiment, one decides whether to give the results
on the unknown parameter $\mu$ in the form of a central confidence interval or
an upper bound. This decision (called `flip-flopping') is based on the data and,
as has been shown by Feldman and Cousins \cite{papiro2}, introduces a bias
that may cause that the intervals cover the true value $\mu$
with a smaller frequency than the stated confidence level.
To solve this and other problems,
they introduce a new ordering principle that unifies the
treatment of central confidence intervals and upper limits.
This is possible because
the Neyman construction of confidence intervals \cite{papiro3} allows the choice
of the ordering principle with which the intervals are constructed. Typical
choices lead to the construction of either central intervals or upper
confidence limits. The choice of Feldman and Cousins gives
intervals that are two-sided or upper limits depending on the result of
the experiment and not on the choice of the experimentalist. These intervals
avoid the undercoverage caused by `flip-flopping' and are
non-empty in all cases. Some variants of their method have been also proposed
\cite{papiro4,papiro5}.

To consider the Feldman-Cousins confidence intervals as an alternative to
standard intervals, in practice one needs to calculate these intervals for 
arbitrary $n_0$ and $b$. The Tables provided in Ref. \cite{papiro2}, for 
$n_0 \leq 20$, $b \leq 15$, are sufficient for small
luminosity/statistics experiments, but for higher luminosities in general $b$
and $n_0$ are larger. One possibility is to extend these
Tables using the same systematic computational method of Ref. \cite{papiro2},
whose speed is not optimized and consumes a lot of time. More
convenient is to develop a program which takes $n_0$ and $b$ as inputs and
gives as output the confidence interval, requiring a minimal number of
calculations. This is what is done here. The program can be used either
directly to compute the
confidence interval for given $n_0$ and $b$ or in conjunction with other
routines. This is especially useful, for instance to calculate expected limits
from rare high energy processes for different values of the center of mass
energy or the collider luminosity, which is the case that we were primarily
interested in.

In the following we introduce a procedure to compute
the Feldman-Cousins intervals in an efficient way for arbitrary $n_0$ and $b$, 
in principle only limited by the machine precision. In Section 2 we review
Neyman's construction of the confidence intervals for a Poisson variable,
emphasizing some points that simplify the numerical calculation. Section 3 is
more technical and devoted to explain in depth how to translate this method
for the computer calculation. In
Section 4 we present our results. The {\tt FORTRAN} implementation of the
algorithm is given in the Appendix. Other implementations in
{\tt C} and {\em Mathematica} \cite{papiro6} (about 100 times slower than the
{\tt FORTRAN} version) can be obtained from the author.

\section{Construction of the confidence intervals}

The probability to observe $n$ events in a Poisson process consisting of signal
events with unknown mean $\mu$ and background events with known mean $b$ is
given by the formula
\begin{equation}
P(n \,|\, \mu;b) = \frac{(\mu+b)^n}{n!} e^{-(\mu+b)} \,,
\label{ec:2.1}
\end{equation}
with $n,\mu,b \geq 0$, and $n$ restricted to integer values. The construction
of the confidence intervals on the unknown variable $\mu$ follows Neyman's
method of the confidence belts.

The first step in this procedure is to construct,
for a fixed value of $b$ and for different values of $\mu$, the
confidence intervals $[n_1(\mu;b),n_2(\mu;b)]$ such that the probability to
obtain a result between $n_1$ and $n_2$ is greater or equal than $\alpha$, the
confidence level (C. L.),
\begin{equation}
P(n \; \epsilon \; [n_1,n_2] \,|\, \mu;b) =
\sum_{n = n_1}^{n_2} P(n \,|\, \mu;b) \geq \alpha \,.
\label{ec:2.2}
\end{equation}
It is worth to note that for the more general case of
a continuous variable $x$ the intervals
$[x_1(\mu;b),x_2(\mu;b)]$ satisfy
$P(n \; \epsilon \; [x_1,x_2] \,|\, \mu;b) = \alpha$.
For a discrete variable $n$ it is not possible to obtain the exact equality,
and to avoid undercoverage it is replaced by the inequality in
Eq. (\ref{ec:2.2}).

The choice of the intervals $[n_1,n_2]$ is not unique, and determines the type
of confidence intervals on $\mu$ that are constructed. The most common choices
are $n_2 = \infty$, $P(n \leq n_1 \,|\, \mu;b) \leq 1-\alpha$, which gives upper
confidence bounds, and $P(n \leq n_1 \,|\, \mu;b) \leq (1-\alpha)/2$,
$P(n \geq n_2 \,|\, \mu;b) \leq (1-\alpha)/2$ which leads to central confidence
intervals. The prescription of Ref. \cite{papiro2} is based on a likelihood
ratio $R$, constructed as follows.
\begin{enumerate}
\item For any values of $b$ and $n$, one considers which value
of $\mu$ would maximize the probability
$P(n \,|\, \mu;b)$ . It is straightforward to
find that for $n \geq 1$, $P(n \,|\, \mu;b)$ considered as a function of $\mu$
grows for $\mu < n-b$, has a maximum at $\mu = n-b$ and decreases for
$\mu > n-b$. As $\mu$ is restricted to lie in the positive real axis, if
$n \geq b$ the maximum is $\mu = n-b$, otherwise the maximum is $\mu = 0$. 
In the case $n = 0$ the maximum is also $\mu = 0$, so we define 
\begin{equation}
\mubest(n;b)=\max\{0,n-b\}
\label{ec:2.3}
\end{equation}
as the value which maximizes $P(n \,|\, \mu;b)$.

\item Then, for any value of $\mu$ we consider the quantity
$R(n;\mu,b)$ defined as
\begin{equation}
R(n;\mu,b) = \frac{P(n \,|\, \mu;b)}{P(n \,|\, \mubest(n;b);b)} \leq 1 \,,
\label{ec:2.4}
\end{equation}
on which the Feldman-Cousins ordering principle is based. To construct the
interval $[n_1(\mu;b),n_2(\mu;b)]$, for each value of $\mu$ (and fixed $b$) one
takes values of $n$ with decreasing $R(n;\mu,b)$,
summing up their probabilities $P(n \,|\, \mu;b)$ until
the total equals or exceeds the C. L. desired $\alpha$. Thus the interval
$[n_1,n_2]$ is the set of values of $n$ necessary to satisfy the
inequality in Eq. (\ref{ec:2.2}), taken with the largest $R(n;\mu,b)$.
\end{enumerate}
The simplicity of this prescription allows a fast computer implementation.
Instead of generating a large table of values for $n$ and taking those
with the largest $R$, we can directly find these values
and add them successively to construct the interval.
(This is difficult to do with the more involved prescriptions of Refs.
\cite{papiro4,papiro5}.)
For this purpose we will examine the behaviour of $R$. If we consider
\begin{equation}
R(x;\mu,b) = \frac{(\mu+b)^x e^{-(\mu+b)}}{(\mubest(x;b)+b)^x
e^{-(\mubest(x;b)+b)}}
\label{ec:2.5}
\end{equation}
as a function of the continous variable $x$, we can look for its maximum.
Let us first consider $\mu \neq 0$, $b \neq 0$. For $x$ sufficiently small,
$\mubest(x;b) = 0$ and $R(x;\mu,b) = (1+\mu/b)^x e^{-\mu}$ is increasing.
For $x \geq b$, $\mubest(x;b) = x-b$ and $R(x;\mu,b) = (e(\mu+b)/x))^x e^{-(\mu
+b)}$ grows for $x < \mu+b$, falls for $x > \mu+b$ and has a local maximum at
$x = \mu+b$, which is then the global maximum. This is
also true for $\mu \neq 0$, $b=0$. 
For $\mu = 0$, $b \neq 0$ and $x < b$, $\mubest(x;b) = 0$ and $R(x;0,b) = 1$,
its maximum possible value. For $x \geq b$, $R(x;0,b) = (b/x)^x e^{x-b}$
decreases with $x$. Hence the maximum is still $x = \mu+b$, although not unique.
The only remaining case with $\mu = 0$, $b = 0$ in which the Poisson
distribution is singular must be treated separately.

\begin{figure}[htb]
\begin{center}
\mbox{
\epsfig{file=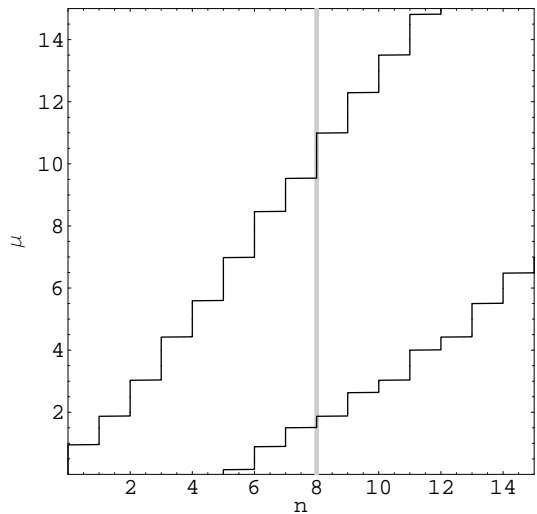,width=7cm}}
\end{center}
\caption{Confidence belt for $b=3$, with $\alpha = 0.9$. The vertical line is
drawn for $\nexp = 8$.
\label{fig:cb1}}
\end{figure}

With this method, and for different values of $\mu$ ($b$ is fixed), one
calculates the confidence intervals $[n_1(\mu;b),n_2(\mu;b)]$ obtaining a 
confidence belt like the one showed in Fig. \ref{fig:cb1} for $b=3$ and a C. L.
of 0.9. To find the confidence interval $[\mu_1,\mu_2]$ for a particular
experimental value $\nexp$,
one draws a vertical line at $n=\nexp$ and finds the maximum and
minimum values of $\mu$ for which the line intersects the confidence belt.
In Fig. \ref{fig:cb1} we observe that $\mu_1$ is the smallest $\mu$
such that $n_2(\mu;b) = \nexp$, whereas $\mu_2$ is the largest
$\mu$ for which $n_1(\mu;b) = \nexp$.

\section{Implementation}

Let us explain how the method described in Section 2 is made suitable
for the evaluation in a computer. For the calculations we use the {\tt FORTRAN}
version of the program compiled with {\tt fort77} under Linux on a Pentium
III-450 (compiled with g77 the program runs about 25\% slower), and for the
plots we also use the {\em Mathematica} version.

The first problem in the practical realization of the Neyman construction is
that, for large $b$, $n$ or $\mu$, the factors of the Poisson
probability formula in
Eq. (\ref{ec:2.1}) can overflow (or underflow) the computer capacity in
intermediate calculations. The
factor $(\mu+b)^n$ may be very large, for instance $54^{178}$ is larger 
the biggest double precision real number in the {\tt FORTRAN} compiler used,
approximately
$10^{308}$. However, the exponential factor in Eq. (\ref{ec:2.1})
compensates for it in the final result. For $\mu+b+n > 230$ we evaluate
{\tt P} using the expression
\begin{equation}
P(n \,|\, \mu;b) = e^{-\mu} \prod_{i=1}^{n} \left( \frac{\mu+b}{i} \right)
e^{-b} \,,
\label{ec:3.1}
\end{equation}
with the product calculated factor by factor. This extends the allowed size of
the parameters of our program, with the disadvantage of a larger computing
time. For $\mu+b+n \leq 230$ the factor $(\mu+b)^n$ is not too large, and can
be directly calculated. In this case the factorials up to $170! \sim 10^{307}$
are calculated at the beginning of the main program and stored in the array
{\tt fact} to save time, whereas for $n > 170$ the expression of the Poisson
formula is divided by $\prod_{i=1}^{n} i$ factor by factor.

The quantity $R$ in Eq. (\ref{ec:2.4}) is a ratio of probabilities and can be
computed without any problem cancelling out the common factors and defining a
function {\tt R}. (Defining $\mubest$ as {\tt DIM(FLOAT(n),b)} instead
of {\tt MAX(0d0,FLOAT(n)-b)} as we do would not have improved the speed
significantly.)

The core of the algorithm is the subroutine {\tt NRANGE}, used to
calculate the confidence intervals $[n_1(\mu;b),n_2(\mu;b)]$ for arbitrary
$\mu$ and $b$. Its arguments are the variables {\tt rmu} ($\mu$), {\tt b} and
the confidence level desired {\tt CL}. The output {\tt n1} and {\tt n2} is
given in a {\tt COMMON} block, together with a variable {\tt CLac}, the C. L.
finally achieved (in general it is greater than {\tt CL}) which is useful for
other purposes.
The discussion in the last Section simplifies the
implementation of the algorithm considerably, because we have found that the
values of $n$ that maximize $R(n;\mu,b)$ concentrate around $\mu+b$. This 
improves the speed by an order of magnitude for large values of the parameters,
since we do not need to calculate a large table $(n,P(n),R(n))$ and sort it.
Instead, we know the maximum {\tt R} is one of the two integers nearest to
$\mu+b$. We begin with {\tt n1=INT(rmu+b)}, {\tt n2=n1+1}. If 
{\tt R(n1,rmu,b)} is larger than {\tt R(n2,rmu,b)} we take {\tt n1}, decrease
{\tt n1} and add {\tt P(n1,rmu,b)} to {\tt CLac}. Otherwise, we take {\tt n2},
increase {\tt n2} and add {\tt P(n2,rmu,b)} to {\tt CLac}. Repeating this until
{\tt CLac} is greater than {\tt CL} and taking into account that {\tt n1} must
be greater than zero we obtain the desired interval $[n_1,n_2]$. The singular
case $\mu = b = 0$ is treated separately. With the {\em Mathematica} version of
this subroutine we can plot confidence belts like that in Fig. \ref{fig:cb1}.

The calculation of the confidence interval $[\mu_1,\mu_2]$ is done using two
funcions {\tt RMU1(n,b,CL)} and {\tt RMU2(n,b,CL)}, where {\tt n} is the
experimental number of events. We discuss them in turn.

\begin{figure}[htb]
\begin{center}
\mbox{
\epsfig{file=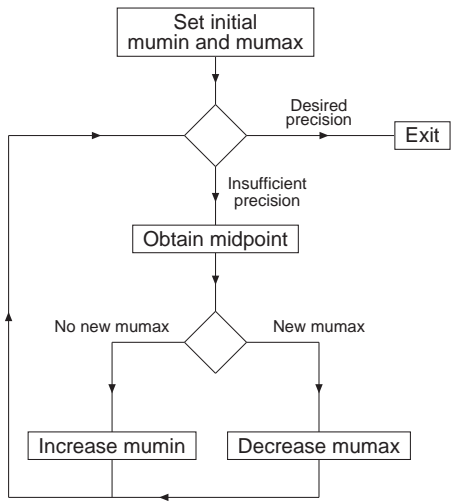,width=6cm}}
\caption{Flux diagram for the calculation of {\tt RMU1}.
\label{fig:df1}}
\end{center}
\end{figure}

As we see in Fig. \ref{fig:cb1}, the lower limit $\mu_1$ is the minimum value
of $\mu$ such that $n_2(\mu;b) = \nexp$. Within our framework, the
calculation is done looking for the minimum {\tt rmu} such that {\tt n2}
calculated with {\tt NRANGE(rmu,b,CL)} equals {\tt n}. The search is done with
the bisection method. Starting with the limits {\tt rmumin=0d0},
{\tt rmumax=FLOAT(n)-b+1d0} ({\tt RMU1} must be between these two values)
we calculate the midpoint of the interval, {\tt rmumed},
and {\tt NRANGE(rmumed,b,CL)}. If {\tt n2} is greater or equal than
{\tt n}, we move {\tt rmumax} to {\tt rmumed}, otherwise we move {\tt rmumin}
to {\tt rmumed}. This is repeated until the length of the interval is smaller
than the desired precision {\tt delta}, which we take as the maximum of 0.01
and 0.0005 times the background $b$. We summarize the algorithm in
Fig. \ref{fig:df1}.

\begin{figure}[htb]
\begin{center}
\mbox{
\epsfig{file=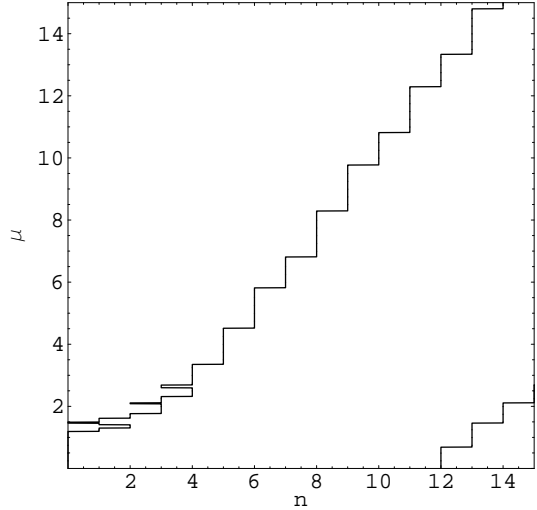,width=7cm}}
\end{center}
\caption{Confidence belt for $b=7$, with $\alpha = 0.95$.
\label{fig:cb2}}
\end{figure}

In principle the calculation of {\tt RMU2} would follow an analogous procedure.
However, in this case we find an extra problem. Except for a few cases with
small $b$, the confidence belt is not as simple as in Fig. \ref{fig:cb1} but
displays a more elaborated structure as can be seen in the example of Fig.
\ref{fig:cb2}. 

The fact that $n_1(\mu;b)$ is not a monotonic function of $\mu$
is due to the discreteness of $n$. This causes that the set of $\mu$ values
for which the vertical line at $n = \nexp$ intersects the belt is not connected
for $\nexp=0-4$. The effect is relevant since the upper
limit $\mu_2$ is defined as the largest $\mu$ for which $n_1(\mu;b) = \nexp$.
Thus a modification of the algorithm is required not to miss the small wedges
in the function. 

For $n \geq b$ the behaviour is as expected and we can use the same algorithm
as for $\mu_1$. We have checked values of $b$ between 0 and 50 and
have found that for $n \geq b$ the function $n_1(\mu;b)$ does not have any
singularity, so in this case we can safely adapt the routine {\tt RMU1}.
We look for the maximum {\tt rmu} such that {\tt n1} calculated with
{\tt NRANGE(rmu,b,CL)} equals {\tt n}. The search is again done with
the bisection method. We start with the initial values
{\tt rmumin=MAX(0d0,FLOAT(n)-b)},
{\tt rmumax=3d0*SQRT(FLOAT(n)+b+1d0)}. (If {\tt rmumax} is not sufficiently
large, we increment it in steps of {\tt SQRT(FLOAT(n)+b+1d0)}.)
We calculate the midpoint of the interval, {\tt rmumed},
and calculate {\tt NRANGE(rmumed,b,CL)}. If {\tt n1} is lower or equal to
{\tt n}, we move {\tt rmumin} to {\tt rmumed}, otherwise we move {\tt rmumax}
to {\tt rmumed}. This is repeated until the length of the interval is smaller
than the desired presision {\tt delta}.

For $n < b$ we sample the interval for possible singularities, which
consumes more time. This is done in three iterations with increasing number of
points. To minimize the length of the interval and optimize
the density of the sampling, we take 
{\tt rmumin=MAX(0d0,FLOAT(n)-b)} and
increase it in steps of 1 while {\tt n1} is lower or equal to {\tt n}. 
As the upper limit we take {\tt rmumax=3d0*SQRT(FLOAT(n)+b+1d0)},
sufficiently high so that the initial
interval contains all the singularities for a C. L. of 0.99 or less. 

In the first step we divide the interval between {\tt rmumin} and {\tt
rmumax} in 10 parts and check if any of the points selected has {\tt n1} lower
or equal to {\tt n}. If
it is so, we change {\tt rmumin} to the largest of them (this is always safe)
and start again with this new {\tt rmumin}. This first sampling with a small
number of points finds wedges like those for $n = 1,3$ in Fig.
\ref{fig:cb2} and saves a lot of computing time. The narrow wedges at
$n=0,2$ require more dense samplings.

If the points calculated have {\tt n1} greater than {\tt n}, we check the
upper half of the interval for singularities. The second iteration divides the
upper half in 20 parts and checks 19 points. If it finds any singular point with
{\tt n1} lower or equal to {\tt n}, it changes {\tt rmumin} and starts again
at the first step. If not, the third iteration divides the upper half in
500 parts. If a singular point is found, it changes {\tt rmumin} and starts the
first step. If not, unless some kind of singular behavior is found (in
which case a fourth sampling with 5000 points is performed) it is assumed that
there do not exist singulariries and {\tt rmumax} is changed.

\begin{figure}[htb]
\begin{center}
\mbox{
\epsfig{file=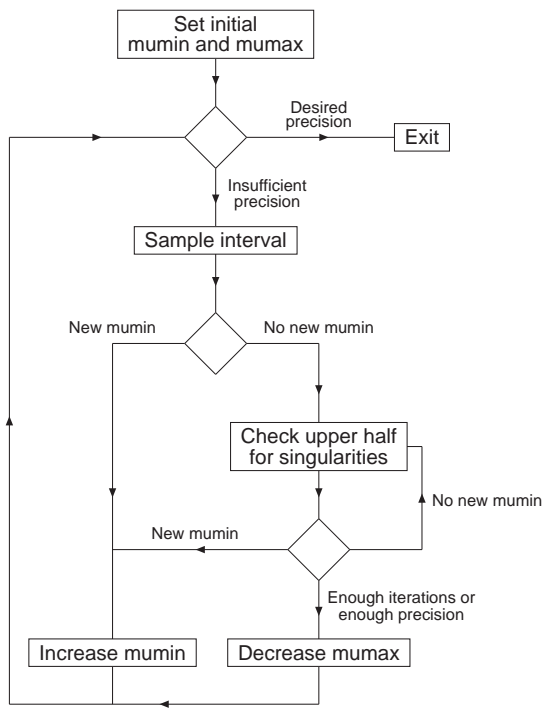,width=7.16cm}}
\caption{Flux diagram for the calculation of {\tt RMU2}.
\label{fig:df2}}
\end{center}
\end{figure}

An additional speed improvement is implemented: if in the second or third
iterations the density of points is sufficiently high (the points are closer
than {\tt stepmin}), the number of points is
decreased and no more iterations are performed. The flux diagram of {\tt RMU2}
is shown in Fig. \ref{fig:df2}.

To simplify changing the parameters of this routine,
the number of iterations and their respective number of
points are stored in the variables {\tt maxit} and {\tt maxdivs}.
 The calculation for $n \geq b$ is done with the
same function with {\tt maxdivs(1)=2} and only the first iteration. 

\begin{figure}[htb]
\begin{center}
\mbox{
\epsfig{file=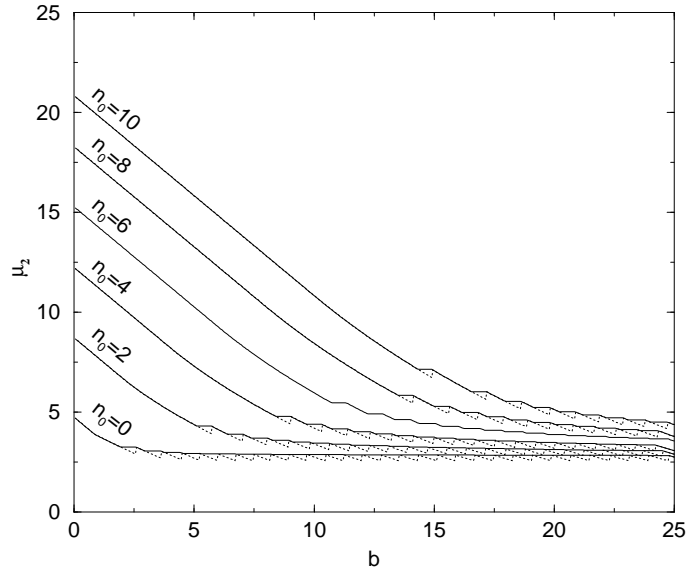,width=9cm}}
\caption{Dependence of $\mu_2$ on the expected background $b$ for fixed $n_0$.
The solid and dotted lines are obtained with and without correction,
respectively.
\label{fig:corr}}
\end{center}
\end{figure}

One may notice that some upper limits obtained with {\tt RMU2} are different
from those quoted in Ref. \cite{papiro2}. This is again a consequence of the
discreteness of $n$. The upper limit $\mu_2$
for $n_0$ fixed is not always a decreasing
function of $b$ (dotted lines in Fig. \ref{fig:corr}.) This behaviour is
corrected in Ref. \cite{papiro2} forcing the function to be nonincreasing,
calculating the upper limit $\mu_2$ from $b = 25$ to $b = 0$ in steps of
$-0.001$. The corrected value is then the maximum of $\mu_2(n_0,x)$ for
$x \geq b$. Some people, however, find this {\em ad hoc}
correction questionable \cite{papiro4}. At any rate we could
also follow this
procedure getting the same values of
Ref. \cite{papiro2} and the solid line in Fig. \ref{fig:corr}. This requires a
very long calculation (25000 different values of $b$), for instance the time
to calculate the $n_0=0$ line is  34 m 27 s.

Of course, to obtain $\mu_2$ for a particular
$b$ it is not necessary to calculate the whole interval $[b,20]$ and it is
enough to consider approximately $[b,b+1]$. For this purpose we use
the function
{\tt RMU2c}. This routine examines the behaviour of {\tt RMU2}
in the interval $[b,b+1]$ and corrects the value if necessary. The adjacent
maxima that can be seen in Fig. \ref{fig:corr} are
found with the simple {\em golden section search} of Ref. \cite{papiro7}.
(Other more sophisticated methods offer no advantage since the function does
not seem to be differentiable at the maximum.)
The initial bracketing of the maximum is very
delicate as can be also seen in this Figure. We do it examining
{\tt RMU2(n,b1,CL)} taking {\tt b1} with increments of 0.1 until {\tt RMU2}
begins to grow, then in increments of 0.05 until it begins to decrease. Then
the {\em golden section} method is applied to find the maximum with a precision
of 0.001. This maximum is then compared to the value at {\tt b} to take
the largest value.

This method again brings a substantial speed improvement over the blind
computation in steps of $-0.001$ in $b$, as we will see in
next Section, Examining the behaviour of $\mu_2$ we have also
found that the correction is not necessary in general for $n > b/2$ and
the function {\tt RMU2} could be used directly. There are however some
exceptions, for instance $n=10$, $b=14$ with a C. L. of 0.95. To be
conservative, we will only use {\tt RMU2} when $n \geq b$.

\section{Results}

To obtain our results we use the same precision that is used in Ref.
\cite{papiro2}, a minimum step {\tt stepmin} of 0.005 in $\mu$ and an accuracy
of 0.01 in the upper and lower limits of the confidence intervals. To calculate
the limits in their Tables II--IX including the singular cases it is enough
to consider in the third iteration {\tt maxdivs(3)=100} for confidence levels
of $68.7\%$, $90\%$ and $95\%$, and {\tt maxdivs(3)=300} for a C. L. of $99\%$.
For better comparison we use {\tt maxdivs(3)=500} as we do in the rest of the
calculations to ensure that all singularities are found.
The running time is summarized in Table \ref{tab:bench1}.

\begin{table}[htb]
\begin{center}
\begin{tabular}{ccc}
C. L. & $t_1$ & $t_2$ \\
68.7\% & 4.1 s & 1 m 9.0 s\\
90\% & 6.1 s & 1 m 28.6 s\\
95\% & 6.8 s & 1 m 36.8 s\\
99\% & 7.9 s & 1 m 57.4 s \\
\end{tabular}
\caption{Time spent in the calculation of the confidence intervals for $n_0
\leq 20$ and $b \leq 15$, with {\tt RMU2} ($t_1$) and with {\tt RMU2c} ($t_2$).
\label{tab:bench1}}
\end{center}
\end{table}

\begin{figure}[htb]
\begin{center}
\mbox{
\epsfig{file=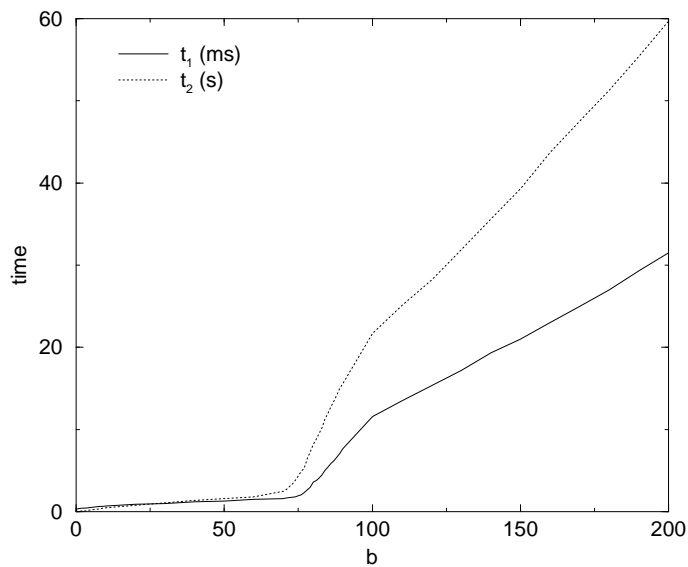,width=9cm}}
\caption{Time required to calculate the upper limit $\mu_2$ with the function
{\tt RMU2} ($t_1$) and with {\tt RMU2c} forcing to look for adjacent maxima
($t_2$) as explained in the text.
\label{fig:bench2}}
\end{center}
\end{figure}

To check if our algorithm in fact handles large numbers efficiently we measure
the time spent to calculate the upper limit $\mu_2$ with {\tt RMU2} for
$n_0 = b$ between 0 and 200, obtaining the solid line in Fig. \ref{fig:bench2}.
We can also use {\tt RMU2c} forcing the program to look for unexisting spurious
maxima in $b$ (and hence also for singularities with {\tt maxdivs(3)=500})
obtaining the dotted line. 

It is amazing to observe that the computing time not only does not grow
quickly with $b$ as it could be expected, but remains almost constant for
$b < 75$. This is achieved with ({\em i\/}) a fast
algorithm to find the singularities if they exist,
 ({\em ii\/}) the optimization of
{\tt NRANGE} to calculate only the data really needed, and ({\em iii\/}) the
calculation of the factorials up to $170!$ at the beginning of the program.
For $b \geq 75$ {\tt rmu+b+FLOAT(n)} is sometimes larger than 230 and
the time required begins to grow linearly with $b$ after the gap
between 75 and 100, as can be observed in the Figure.

\vspace{1cm}
\noindent
{\Large \bf Acknowledgements}

\vspace{0.4cm} \noindent
I thank F. del Aguila for discussions and for a critical reading of the
manuscript.
This work was partially supported by CICYT under contract AEN96--1672 and by
the Junta de Andaluc\'{\i}a, FQM101.

\appendix
\section{Appendix}
\begin{verbatim}
      PROGRAM DEMO
      IMPLICIT REAL*8 (A-H,O-Z)
      DIMENSION FACT(0:170)
      COMMON /range/ n1,n2,CLac
      COMMON /factorial/ FACT
      DATA CL /0.99d0/

      FACT(0)=1d0                           ! FACT initialization
      DO i=1,170
        FACT(i)=FACT(i-1)*DFLOAT(i)
      ENDDO
            
C     CALCULATION OF TABLES VIII, IX OF REF[2]

      DO b=0d0,4d0,0.5d0
        DO n=0,20
          PRINT 100,n,b,RMU1(n,b,CL),RMU2c(n,b,CL)
        ENDDO
      ENDDO
      DO b=5d0,15d0,1d0
        DO n=0,20
          PRINT 100,n,b,RMU1(n,b,CL),RMU2c(n,b,CL)
        ENDDO
      ENDDO
      STOP

100   FORMAT ('n = ',I2,', b = ',F4.1,' -> [',F6.2,',',F6.2,']')
      END


      DOUBLE PRECISION FUNCTION P(n,rmu,b)
      IMPLICIT REAL*8 (A-H,O-Z)
      DIMENSION FACT(0:170)
      COMMON /factorial/ FACT
      IF ((rmu .EQ. 0d0) .AND. (b .EQ. 0d0)) THEN    ! These lines
        P=0                                          ! are not
        IF (n .EQ. 0) P=1                            ! needed if P is
        RETURN                                       ! only called
      ENDIF                                          ! from NRANGE
      IF (rmu+b+FLOAT(n) .LE. 230d0) THEN
        P=(rmu+b)**n*EXP(-rmu-b)
        IF (n .LE. 170) THEN
          P=P/FACT(n)
        ELSE
          P=P/FACT(170)    ! This is not normally used because for
          DO i=171,n       ! n > 170 rmu+b+FLOAT(n) will be larger
            P=P/FLOAT(i)   ! than 230
          ENDDO
        ENDIF
      ELSE
        P=EXP(-rmu)
        DO i=1,n
          P=P*(rmu+b)/FLOAT(i)
        ENDDO
        P=P*EXP(-b)
      ENDIF
      RETURN
      END
      
      DOUBLE PRECISION FUNCTION R(n,rmu,b)
      IMPLICIT REAL*8 (A-H,O-Z)
      IF (n .LT. 0) THEN
        R=0d0
        RETURN
      ENDIF
      R=EXP(MAX(0d0,FLOAT(n)-b)-rmu)
      IF (n .GT. 0) R=R*((rmu+b)/(MAX(0d0,FLOAT(n)-b)+b))**n
      RETURN
      END


      SUBROUTINE NRANGE(rmu,b,CL)
      IMPLICIT REAL*8 (A-H,O-Z)
      COMMON /range/ n1,n2,CLac  ! CLac for future use
      IF ((rmu .EQ. 0d0) .AND. (b .EQ. 0d0)) THEN
        n1=0d0                   ! Special case
        n2=0d0
        RETURN
      ENDIF
      n1=INT(rmu+b)              ! The maximum R is between
      n2=n1+1                    ! these values
      r1=R(n1,rmu,b)
      r2=R(n2,rmu,b)
      CLac=0d0
      DO WHILE ((CLac .LT. CL) .AND. (n1 .GE. 0))
        IF (r1 .GT. r2) THEN
          CLac=CLac+P(n1,rmu,b)
          n1=n1-1
          r1=R(n1,rmu,b)
        ELSE
          CLac=CLac+P(n2,rmu,b)
          n2=n2+1
          r2=R(n2,rmu,b)
        ENDIF
      ENDDO
      DO WHILE (CLac .LT. CL)    ! No need to calculate R
        CLac=CLac+P(n2,rmu,b)
        n2=n2+1
      ENDDO
      n1=n1+1
      n2=n2-1
      RETURN
      END


      DOUBLE PRECISION FUNCTION RMU1(n,b,CL)
      IMPLICIT REAL*8 (A-H,O-Z)
      COMMON /range/ n1,n2,CLac
      CALL NRANGE(0d0,b,CL)
      IF (n2 .GE. n) THEN
        RMU1=0d0                                ! Special case
        RETURN
      ENDIF
      rmumin=0d0
      rmumax=FLOAT(n)-b+1d0
      delta=MAX(0.01d0,0.0005d0*b)
      DO WHILE ((rmumax-rmumin) .GE. delta)     ! Bisection method
        rmumed=(rmumin+rmumax)/2d0
        CALL NRANGE(rmumed,b,CL)
        IF (n2 .GE. n) THEN
          rmumax=rmumed
        ELSE
          rmumin=rmumed
        ENDIF
      ENDDO
      RMU1=(rmumin+rmumax)/2d0
      RETURN
      END


      DOUBLE PRECISION FUNCTION RMU2(n,b,CL)
      IMPLICIT REAL*8 (A-H,O-Z)
      DIMENSION maxdivs(4)
      LOGICAL safe,safenow,sing,changemin
      COMMON /range/ n1,n2,CLac
      DATA maxdivs /10,20,500,5000/
      maxit=4
      stepmin=0.005d0
      rmumin=MAX(0d0,FLOAT(n)-b)
      CALL NRANGE(rmumin,b,CL)
      IF (FLOAT(n) .LT. b) THEN
        DO WHILE (n1 .LE. n)          ! Take lower limit
          rmumin=rmumin+1d0           ! as high as possible
          CALL NRANGE(rmumin,b,CL)
        ENDDO
        rmumin=rmumin-1d0
        safe=.FALSE.
      ELSE
        maxdivs(1)=2                  ! Use bisection method when
        safe=.TRUE.                   ! there aren't sing.
      ENDIF
      rmumax=3d0*SQRT(FLOAT(n)+b+1d0) ! Large enough for most purposes
      CALL NRANGE(rmumax,b,CL)
      DO WHILE (n1 .LE. n)            ! If not, increase it
        rmumax=rmumax+SQRT(FLOAT(n)+b+1d0)
        CALL NRANGE(rmumax,b,CL)
      ENDDO
      delta=MAX(0.01d0,0.0005d0*b)
      DO WHILE ((rmumax-rmumin) .GE. delta)
        step=(rmumax-rmumin)/FLOAT(maxdivs(1))
        rmumin2=rmumin
        DO i=1,maxdivs(1)-1
          CALL NRANGE(rmumin+FLOAT(i)*step,b,CL)
          IF (n1 .LE. n) rmumin2=rmumin+FLOAT(i)*step
        ENDDO
        IF (rmumin2 .GT. rmumin) THEN
          rmumin=rmumin2              ! New rmumin -> change it
        ELSE
          safenow=safe                ! Have to look for singularities
          sing=.FALSE.                ! if they may exist
          changemin=.FALSE.
          it=2
          DO WHILE ((safenow .EQ. .FALSE.) .AND. (it .LE. maxit))
            ndivs=maxdivs(it)
            step=(rmumax-rmumin)/FLOAT(2*ndivs)
            IF (step .LT. stepmin) THEN                ! step is small
              ndivs=INT((rmumax-rmumin)/(2*stepmin))+1 ! enough and this
              step=(rmumax-rmumin)/FLOAT(2*ndivs)      ! will be the
              safenow=.TRUE.                           ! last iteration
            ENDIF
            CALL NRANGE((rmumin+rmumax)/2d0,b,CL)
            n_prev=n1
            DO i=1,ndivs-1
              CALL NRANGE(rmumin+FLOAT(i+ndivs)*step,b,CL)
              IF (n1 .LE. n) THEN
                rmumin2=rmumin+FLOAT(i+ndivs)*step     ! New rmumin
                changemin=.TRUE.
              ENDIF
              IF (n1 .LT. n_prev) sing=.TRUE.
              n_prev=n1
            ENDDO
            IF (changemin .EQ. .TRUE.) THEN
              rmumin=rmumin2                    ! Change rmumin
              safenow=.TRUE.                    ! Exit loop
            ELSE 
              IF ((sing .EQ. .FALSE.) .AND. (it .EQ. maxit-1)) THEN
                safenow=.TRUE.                  ! Enough iterations
              ENDIF
            ENDIF
            it=it+1                             ! Next iteration
          ENDDO
          IF (changemin .EQ. .FALSE.) rmumax=(rmumin+rmumax)/2d0
        ENDIF
      ENDDO
      RMU2=(rmumin+rmumax)/2d0
      RETURN
      END


      DOUBLE PRECISION FUNCTION RMU2c(n,b,CL)
      IMPLICIT REAL*8 (A-H,O-Z)
      LOGICAL sing
      PARAMETER (R=0.61803399d0,C=1d0-R)
      RMU2c=RMU2(n,b,CL)
      IF (n .GE. INT(b)) RETURN        ! Do not need correction
      step1=0.1d0                      ! Go downhill in steps of 0.1
      step2=0.05d0                     ! Go uphill in steps 0f 0.05
      deltab=0.001d0                   ! Final precision in b
      b1max=b+1d0                      ! Look for maximum up to b+1
      sing=.FALSE.

C     Bracket the maximum, if any

      b1=b                             
      a1=RMU2c
      DO WHILE ((b1 .LE. b1max) .AND. (sing .EQ. .FALSE.)) ! Go downhill
        a_next=RMU2(n,b1+step1,CL)
        IF (a_next .GT. a1) THEN
          sing=.TRUE.
        ELSE
          a1=a_next
          b1=b1+step1
        ENDIF
      ENDDO
      IF (sing .EQ. .FALSE.) RETURN    ! RMU2 is always decreasing
      b2=b1+step1-step2
      a2=a_next
      DO WHILE (a_next .GE. a2)        ! Go uphill
        a2=a_next
        b2=b2+step2
        a_next=RMU2(n,b2+step2,CL)
      ENDDO
      b4=b2+step2
      a4=a_next
      IF (RMU2c .GT. a2+0.05d0) RETURN  ! This maximum will not be larger
      IF (b4-b2 .GT. b2-b1) THEN          
        b3=b2+C*(b4-b2)
        a3=RMU2(n,b3,CL)
      ELSE
        b3=b2
        a3=a2
        b2=b3-C*(b3-b1)
        a2=RMU2(n,b2,CL)
      ENDIF

C     Find the maximum

      DO WHILE (b4-b1 .GE. deltab)
        IF (a3 .GT. a2) THEN
          b1=b2
          b2=b3
          b3=R*b2+C*b4
          a1=a2
          a2=a3
          a3=RMU2(n,b3,CL)
        ELSE
          b4=b3
          b3=b2
          b2=R*b3+C*b1
          a4=a3
          a3=a2
          a2=RMU2(n,b2,CL)
        ENDIF
      ENDDO
      RMU2c=MAX(RMU2c,a2,a3)
      RETURN
      END
\end{verbatim}


\begin{thebibliography}{99}
\bibitem{papiro1}
K. Eitel and B. Zeitnitz, hep-ex/9809007;
B. Zeitnitz, talk presented at Neutrino '98. See \\
{\tt http://www-ik1.fzk.de/www/karmen/karmen.e.html}
\bibitem{papiro2}
G. J. Feldman y R. D. Cousins, Phys. Rev. {\bf D57}, 3873 (1998)
\bibitem{papiro3}
J. Neyman, Philos. Trans. R. Soc. London {\bf A236}, 333 (1937). Reprinted
in {\em A selection of Early Statistical Papers on J. Neyman}, University
of California Press, Berkeley 1967
\bibitem{papiro4}
C. Giunti, Phys. Rev. {\bf D59}, 053001 (1999)
\bibitem{papiro5}
B. P. Roe and M. B. Woodroofe, Phys. Rev. {\bf D60}, 053009 (1999)
\bibitem{papiro6}
S. Wolfram, {\em Mathematica, a System for Doing Mathematics by Computer},
Addison-Wesley Publishing Company, Redwood City, California, 1988
\bibitem{papiro7}
W. H. Press, B. P. Flannery, S. A. Teukolsky y W. T. Vetterling,
{\em Numerical Recipes}, Cambridge University Press 1986
\end{thebibliography}
\end{document}